\documentclass[useAMS,usenatbib,usegraphicx]{mn2e}

\usepackage{times}
\usepackage{multicol}

\newcommand{\AIPS}{{$\cal AIPS\/$}}

\def\gs{\mathrel{\raise0.35ex\hbox{$\scriptstyle >$}\kern-0.6em
\lower0.40ex\hbox{{$\scriptstyle \sim$}}}}
\def\ls{\mathrel{\raise0.35ex\hbox{$\scriptstyle <$}\kern-0.6em
\lower0.40ex\hbox{{$\scriptstyle \sim$}}}}

\title[Interferometric imaging of a forming, massive elliptical]
      {Interferometric imaging of the high-redshift radio galaxy, 4C\,60.07:
       An SMA, {\em Spitzer} and VLA study reveals a binary AGN/starburst}
      \author[Ivison et al.]
             {R.\,J.\ Ivison,$^{\! 1,2}$
              G.\,E.\ Morrison,$^{\! 3,4}$ A.\,D.\ Biggs,$^{\! 1}$
              Ian Smail,$^{\! 5}$ S.\,P.\ Willner,$^{\! 6}$ 
              M.\,A.\ Gurwell,$^{\! 6}$ \and T.\,R.\ Greve,$^{\! 7}$
              J.\,A.\ Stevens$^{8}$ and M.\,L.\,N.\ Ashby$^{6}$\\
              \vspace*{1mm}\\
      $^1$UK Astronomy Technology Centre, Royal Observatory, Blackford Hill,
          Edinburgh EH9 3HJ\\
      $^2$Scottish Universities Physics Alliance,
          Institute for Astronomy, University of Edinburgh, Royal Observatory,
          Blackford Hill, Edinburgh EH9 3HJ\\
      $^3$Institute for Astronomy, University of Hawaii, Honolulu, HI 96822, USA\\
      $^4$Canada-France-Hawaii Telescope, Kamuela, HI 96743, USA\\
      $^5$Institute for Computational Cosmology, University of Durham,
          South Road, Durham DH1 3LE\\
      $^6$Harvard-Smithsonian Center for Astrophysics, 60 Garden Street, Cambridge,
          MA 02138, USA\\
      $^7$Max-Planck Institute for Astronomy, Heidelberg, Germany\\
      $^8$Centre for Astrophysics Research, University of Hertfordshire,
          College Lane, Hatfield AL10 9AB
}

\date{Accepted 2008 August 7.
      Received 2008 August 5; in original form 2008 June 19}

\pagerange{000--000}

\begin{document}

\maketitle

\begin{abstract}
High-resolution submillimetre (submm) imaging of the high-redshift
radio galaxy (HzRG), 4C\,60.07, at $z=3.8$, has revealed two dusty
components of roughly equal integrated flux. {\em Spitzer} imaging
shows that one of these components (`B') is coincident with an
extremely red active galactic nucleus (AGN), offset by $\sim$4\,arcsec
($\sim$30\,kpc) from the HzRG core. The other submm component (`A') --
resolved by our synthesised beam and devoid of emission at
3.6--8.0-$\mu$m -- lies between `B' and the HzRG core. Since the radio
galaxy was discovered via its extremely young, steep-spectrum radio
lobes and the creation of these lobes was likely triggered by the
interaction, we argue that we are witnessing an early-stage merger,
prior to its eventual equilibrium state. The interaction is between
the host galaxy of an actively-fueled black hole (BH), and a gas-rich
starburst/AGN (`B') marked by the compact submm component and
coincident with broad CO(4--3) emission. The second submm component
(`A') is a plume of cold, dusty gas, associated with a narrow
($\sim$150\,km\,s$^{-1}$) CO feature, and may represent a short-lived
tidal structure. It has been claimed that HzRGs and submm-selected
galaxies (SMGs) differ only in the activity of their AGNs, but such
complex submm morphologies are seen only rarely amongst SMGs, which
are usually older, more relaxed systems. Our study has important
implications: where a galaxy's gas reservoir is not aligned with its
central BH, CO may be an unreliable probe of dynamical mass, affecting
work on the co-assembly of BHs and host spheroids. Our data support
the picture wherein close binary AGN are induced by mergers. They also
raise the possibility that some supposedly jet-induced starbursts may
have formed co-evally (yet independently of) the radio jets, both
triggered by the same interaction.  Finally, we note that the HzRG
host would have gone unnoticed without its jets and its companion, so
there may be many other unseen BHs at high redshift, lost in the sea
of $\sim5\times 10^8$ similarly bright IRAC sources -- sufficiently
massive to drive a $>$10$^{27}$-W radio source, yet practically
invisible unless actively fueled.
\end{abstract}


\section{Introduction}

Radio galaxies are believed to host actively-fueled, spinning BHs
which power their immense radio luminosities and fashion their
characteristic double lobes \citep*{fr74,rees78,bp82,mccarthy93}.
High-redshift examples of the phenomenon (HzRGs) have often been
identified as ultra-steep-spectrum (USS) emitters in radio
surveys\footnote{USS sources, with $\alpha<-1$ where $S_{\nu}\propto
\nu^{\alpha}$, were often found to lack identifications on
photographic plates \citep*[e.g.][]{tielens79}.}
\citep[e.g.][]{bornancini07} and are thus selected during their
extreme youth \citep*[$\le$10\,Myr --][]{br99}. Nowadays, a wealth of
unequivocal evidence also links HzRGs with very massive galaxies in
the early Universe. Near- and mid-infrared (-IR) observations show
that HzRGs are associated with the most massive stellar populations at
any given redshift \citep*{blr97, seymour07}.  Direct evidence of vast
reservoirs of atomic and molecular hydrogen has also been established,
via observations of strong H\,{\sc i} absorption against luminous,
morphologically complex Ly$\alpha$ halos -- often extending
100--200\,kpc from the central radio galaxy -- and via detections of
CO and dust \citep*{hm93, papadopoulos00, archibald01, reuland03,
reuland04, debreuck03, debreuck05, klamer05, villarmartin06}.

If HzRGs pinpoint the most massive galaxies out to the highest
redshifts then we expect these systems to signpost the most massive
peaks in the primordial Universe and to evolve into today's brightest
galaxies -- cD-type ellipticals\footnote{Defined as being centrally
located in clusters and very much larger and brighter than other
galaxies in the cluster, the {\em c} being analogous to the notation
for supergiant stars in stellar spectroscopy \citep{mms64}.}. In
accord with these expectations, the environments of HzRGs are
over-dense in a rich variety of galaxy types, including SMGs. The
first example, still the most striking, is the field of 4C\,41.17 at
$z=\rm 3.8$ \citep{ivison00, greve07}, and \citet{stevens03} presented
low-resolution submm imaging of a further six HzRGs. Some of these
displayed evidence of extended ($\sim$5--20\,arcsec,
$\sim$30--150\,kpc) dust emission \citep[and UV emission
--][]{hatch08}, suggesting that starbursts in HzRGs differ from the
compact events seen in local ultraluminous IR galaxies (ULIRGs) and,
indeed, in the general $z\sim\rm 1-3$ SMG population as probed by
high-resolution CO and submm--radio continuum imaging
\citep[$\ls$0.5\,arcsec, or $\ls$4\,kpc --][]{chapman04, tacconi06,
tacconi08, bi08, younger08b}, i.e.\ that the mechanism for the
formation of these very massive galaxies may be fundamentally
different to that of $\gs$L$_{\star}$ galaxies forming from SMGs
\citep{smail04}.

Probing the size and morphology of the rest-frame far-IR emission in
HzRGs is thus a key piece in the puzzle of galaxy formation. With the
$\sim$14-arcsec ({\sc fwhm}) spatial resolution usually available, we
have been unable to distinguish between entirely different modes for
the formation of the most massive galaxies. One could imagine HzRG
host galaxies assembling via a massive, widespread burst with a low
overall star-formation density, or within multiple gas-rich
sub-components destined to merge. Evidence from mm interferometric
imaging has been enlightening, though the need for imaging on a range
of spatial scales is often evident: \citet{debreuck05} found evidence
for two gas components in the 4C\,41.17 system whilst
\citet{papadopoulos00} argued for a galaxy-wide starburst in another
HzRG, 4C\,60.07, with gas being consumed on a scale of $\sim$30\,kpc,
and for a gas reservoir more commensurate with local ULIRGs in
another, 8C\,1909+722.

4C\,60.07 is a powerful double-lobed FR\,{\sc ii} radio galaxy -- one
of the brightest of the HzRGs observed by \citet{archibald01} and
\citet{reuland04} in the submm waveband. The radio spectrum of
4C\,60.07 is unremarkable at low frequencies, with a spectral index,
$\alpha=-0.9$ between 38 and 178\,MHz (where $S_{\nu} \propto
\nu^{\alpha}$), but its spectrum steepens at higher frequencies,
reaching $\alpha=-1.6$ between 4.9 and 15\,GHz, hence its selection as
an USS radio emitter and its eventual identification with an $R_{\rm
Vega}\sim 23.2$ galaxy at $z=\rm 3.7887\pm 0.0007$ \citep{chambers96,
rottgering97, reuland07}.

In this paper we present submm interferometry of 4C\,60.07, utilising
the Submillimetre Array\footnote{The Submillimeter Array is a joint
project between the Smithsonian Astrophysical Observatory and the
Academia Sinica Institute of Astronomy and Astrophysics and is funded
by the Smithsonian Institution and the Academia Sinica.} \citep[SMA
--][]{ho04} at a wavelength of 890\,$\mu$m with a spatial resolution
of $\sim$2\,arcsec and high astrometric precision -- the first
high-resolution submm continuum image of an unmistakably massive
galaxy at $z\sim\rm 4$, sensitive to dust on the scales predicted by
existing single-dish imaging. The primary goal was to determine the
size and nature of the emission. Is it diffuse, or is it due to
multiple discrete, compact objects whose total flux density can
account for the high $S_{\rm 850\mu m}$ seen by SCUBA?

In the next section we describe an extensive set of observations and
the associated analysis. Presentation of the reduced images follows in
\S\ref{discussion}, with our interpretation of those data and
discussion of their implications in \S\ref{model} and
\S\ref{conclusions}.

\section{Observations}
\label{observations}

\subsection{SMA}
\label{sma}

\begin{figure}
\begin{center}
\includegraphics[scale=0.35]{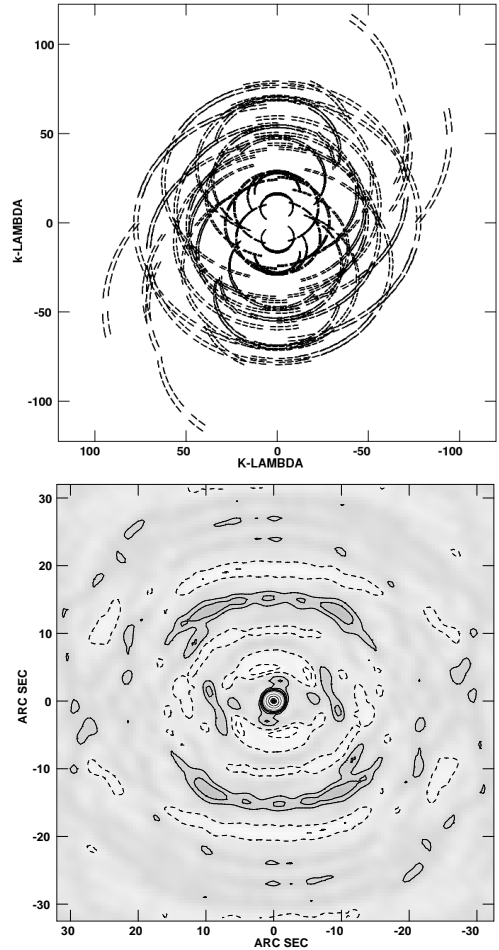}
\includegraphics[scale=0.35]{fig1b.eps}
\caption{{\em Top:} Coverage in the $uv$ plane of SMA data acquired on
2007 March 02 and October 23. {\em Bottom:} Greyscale representation
of the dirty beam, with contours at $-10$, $-5$, 5, 10, 15, 20, 40, 60
and 80 per cent of the peak.}
\label{uv}
\end{center}
\end{figure}

Data were obtained\footnote{Additional data were taken in poor weather
conditions during 2007 February 20 -- see Table~\ref{obs-tab} -- but
were not used here.}  on Mauna Kea, Hawaii, during 2007 March and
October using the SMA's ``compact'' configuration (baselines as long
as 113\,m during March and 70\,m during October -- see Fig.~\ref{uv})
and the 345-GHz receivers, with 2\,GHz of bandwidth in each of the
upper and lower sidebands, as outlined in Table~\ref{obs-tab}. The
pointing centre was R.A.\ 05:12:54.80, Dec.\ +60:30:51.7 J2000 and
observations of 4C\,60.07 were interspersed with $\sim$5-min scans of
the phase calibrator, 4C\,50.11, a 2.3-Jy source located 14$^{\circ}$
from the target. The October data also included observations of
8C\,0716+714 (1\,Jy, 17$^{\circ}$) -- to calibrate those data taken
when the target was setting -- and NVSS\,J044923+633208 (0.3\,Jy,
4$^{\circ}$) to check the quality of the phase referencing. The
position of NVSS\,J044923+633208 was found to be accurate to
$\ls$0.1\,arcsec.

All calibration was carried out using {\sc mir}, a set of {\sc
idl}-based routines specifically developed for data from the SMA. The
main steps consisted of flagging noisy or corrupted data, converting
the data to units of flux density using antenna system temperatures,
determining the bandpass (with observations of 3C\,454.3 or 3C\,279)
and, finally, calibrating the antenna complex gains (phase and
amplitude) and interpolating the calibrator solutions onto the
target. The absolute flux scale is estimated to be accurate to 10 per
cent. The data were then read into the Astronomical Image Processing
Software (\AIPS) package where they were imaged.

The duration of our October track was 7.35\,hr, with around 1\,mm of
precipitable water vapour (pwv). The resulting noise level was
1.4\,mJy\,beam$^{-1}$, with a [2.4 $\times$ 2.2]\,arcsec {\sc fwhm}
synthesised beam for naturally weighted $uv$ data.  The SMA Beam
Calculator and Sensitivity Estimator\footnote{\tt
http://sma1.sma.hawaii.edu/beamcalc.html} had predicted a synthesised
beam profile of [2.2 $\times$ 2.1]\,arcsec {\sc fwhm} and an r.m.s.\
noise level of 1.41\,mJy\,beam$^{-1}$, for a 7.5-hr track with a
20$^{\circ}$ minimum elevation and 1\,mm of pwv, and so appears to
function adequately.

We experimented with several permutations of data. All combinations
resulted in images with the same basic characteristics.  Although
combining all the data yielded the lowest noise,
1.19\,mJy\,beam$^{-1}$, we opted to use only those data taken in
genuine `submm weather' ($\tau_{\rm 225GHz}<0.1$), on October 23 and
March 02, since experience suggests only these produce adequate
representations of the submm sky. The resulting noise level was
1.38\,mJy\,beam$^{-1}$, before {\sc clean}ing, with a synthesised beam
measuring [2.25 $\times$ 2.07]\,arcsec {\sc fwhm} with the major axis
at position angle (PA), 78$^{\circ}$. Employing 100 interations of the
{\sc clean} algorithm, with a gain of 0.1, in a region of radius
45\,arcsec centred on the radio galaxy (nine times the area contained
within the SMA's primary beam), resulted in a 20 per cent reduction in
the measured noise. The source structure and flux density were
unaffected.

\begin{table}
\begin{center}
\caption{Observations of 4C\,60.07 with the SMA.}
\begin{tabular}{lccc} \\ \hline
Date  & Track length & Conditions & Comment\\
(2007)& /min         & $\tau_{\rm 225GHz}$ &\\ \hline
February 20 & 216 & $0.167\pm0.012$& Unsuitable \\ 
March 02 & 257 & $0.063\pm0.012$& Adequate\\
March 03 & 268 & $0.108\pm0.008$& Marginal\\
October 23 & 441 & $0.053\pm0.004$& Excellent\\ \hline
\end{tabular}
\label{obs-tab}
\end{center}
\end{table}

\subsection{\em Spitzer}
\label{spitzer}

4C\,60.07 was observed with {\em Spitzer}\footnote{This work is based
in part on observations made with the {\em Spitzer Space Telescope},
which is operated by the Jet Propulsion Laboratory, California
Institute of Technology under a contract with NASA.  Support for this
work was provided by NASA through an award issued by JPL/Caltech.}
\citep{werner04} in all the broadband filters available in the
Multiband Imaging Photometer for {\em Spitzer} \citep[MIPS
--][]{rieke04} and Infrared Array Camera \citep[IRAC --][]{fazio04}
bands, 3.6--8.0 and 24--160\,$\mu$m, respectively, in 2004 October,
i.e.\ during the cold portion of the mission \citep[programme
identification: 3329 --][]{seymour07}.

The IRAC data consist of a short sequence of four dithered 30-s frames
centred on 4C\,60.07 in each of the four bands.  We eliminated
scattered light, column pulldown, residual images and multiplexer
bleed from the individual exposures by hand, then mosaiced these
cosmetically superior frames using {\sc irac}proc \citep{schuster06}
with 0.86-arcsec pixels (half the area of the native IRAC pixels).

The Infrared Spectrograph \citep[IRS --][]{houck04} also took two 61-s
exposures at 16\,$\mu$m in the 4C\,60.07 field, using `peak-up
imaging' mode, prior to it being offered to the community. We use the
image reduced by \citet{seymour07} here.

The reduction of the 24-$\mu$m data was based on version S16.1.0 of
the MIPS pipeline. We used an object-masked median stack of all the
MIPS exposures of 4C\,60.07 to compensate for structure arising from
residual images. The remaining structure in the backgrounds was
eliminated using {\sc imsurfit} in {\sc iraf} and the resulting
mosaics were created with 1.2-arcsec pixels.

The IRAC, IRS and MIPS data are aligned to less than 0.5\,arcsec,
conservatively.

For IRAC and MIPS, photometric measurements were made using 4.9- and
7.0-arcsec-diameter circular apertures, respectively, correcting to
total flux via the standard aperture correction\footnote{IRAC Data
  Handbook; http://ssc.spitzer.caltech.edu/irac/dh/}. The flux did not
vary significantly for apertures ranging from 6 to 9\,arcsec. At all
wavelengths, apertures were centred on the 4.5-$\mu$m centroid.

\subsection{Optical and near-IR observations}
\label{optir}

An $I$-band image was obtained during 2000 November 01 using the 4.2-m
William Herschel Telescope's (WHT\footnote{Based on observations made
with the WHT operated on the island of La Palma by the Isaac Newton
Group in the Spanish Observatorio del Roque de los Muchachos of the
Instituto de Astrofisica de Canarias}) Mosaic Camera, which comprises
two EEV 4096\,$\times$\,2048 detectors with 0.236-arcsec pixels
\citep{tullock00}. The 3-$\sigma$ limiting magnitude in a
4.9-arcsec-diameter aperture is $I_{\rm Vega}=24.8$.

We also utilise the $K'$ imaging of \citet{vb98}, a 1-hr exposure
taken using NIRC \citep{ms94} on the 10-m Keck {\sc i} telescope
during 0.65-arcsec seeing, with a 3-$\sigma$ limiting magnitude in a
4.9-arcsec-diameter aperture of $K'_{\rm Vega}=22.0$.

\subsection{Radio observations}
\label{radio}

Pseudo-continuum data were obtained at 1.4\,GHz during 2001 January
using the NRAO's Very Large Array (VLA) in its A configuration
(programmes AD432 and AI84). The raw data were flagged, calibrated and
imaged following the technique described by \citet{bi06} and the
resulting images have a synthesised beam size of [1.4 $\times$
1.1]\,arcsec at a PA of 26$^{\circ}$.

We obtained still higher-resolution continuum images, with 100-MHz
bandwidths centred at 4.7 and 8.2\,GHz (programme AC374) and
synthesised beam sizes of $\sim$0.5 and $\sim$0.3\,arcsec {\sc fwhm}
from the National Radio Astronomy Observatory's\footnote{The National
Radio Astronomy Observatory is operated by Associated Universities
Inc., under a cooperative agreement with the National Science
Foundation.} (NRAO's) Data Archive System\footnote{\tt
https://archive.nrao.edu}.

We also reduced the 24-GHz continuum data taken by \citet{gip04} to
explore CO $J=1-0$ emission from 4C\,60.07 for the velocity range of
the broad CO(4--3) component found by \citet{papadopoulos00}: $+455\pm
312$\,km\,s$^{-1}$. These were obtained between 2001 October and 2003
March and include five, two and two tracks of data in the D, C and DnC
configurations, respectively (programmes AI88, AI93 and
AI104). Combining these data with a natural weighting scheme ({\sc
robust} = 5) yielded a [2.6 $\times$ 2.4]-arcsec synthesised beam
({\sc fwhm}), at PA 102$^{\circ}$, and $\sigma=
11.9\,\mu$Jy\,beam$^{-1}$ after 100 iterations of the {\sc clean}
algorithm in \AIPS\ {\sc imagr} with a gain of 0.1 -- more than twice
as deep as the maps presented by \citet{gip04} who used only the
highest resolution data.

We used several methods to quantify the CO $J=1-0$ emission. First, we
aped the approach of \citet{gip04}: we imaged the two 50-MHz-wide
intermediate frequencies (IFs), representing `continuum only' (IF1,
$\sigma = 15.7\,\mu$Jy\,beam$^{-1}$) and `continuum plus line' (IF2,
$\sigma = 15.9\,\mu$Jy\,beam$^{-1}$), then subtracted IF1 {\sc clean}
components from the $uv$ database for IF2, after checking the position
and integrated flux density of the brightest lobe in IF1 and IF2. The
dirty image made using the continuum-subtracted data has
$\sigma=16.0\,\mu$Jy\,beam$^{-1}$.  For our second method, we
subtracted a dirty image of IF1 from one of IF2, yielding a $\sim\sqrt
2\times$ noisier map than the first method ($\sigma =
23.5\,\mu$Jy\,beam$^{-1}$). For our third method, we repeated the
second method for each of the nine observational epochs, tapering the
data to produce a $\sim$5-arcsec ({\sc fwhm}) beam. The noise-weighted
combination of these nine images yielded $\sigma =
30.0\,\mu$Jy\,beam$^{-1}$. The three methods gave a consistent picture
in which the peak flux density due to broad-line CO $J=1-0$ emission
centred on any of the prominent IR or radio components can be no
greater than 0.05\,mJy\,beam$^{-1}$, and no greater than 0.12\,mJy in
total (both 3\,$\sigma$). 

\subsection{Submm observations}
\label{submm}

Observations using the Submm Common-User Bolometer Array \citep[SCUBA
  --][]{holland99} on the 15-m James Clerk Maxwell Telescope (JCMT) at
a wavelength of 850\,$\mu$m were described by \citet{stevens03}. We
have returned to those data, and the 450-$\mu$m data obtained
simultaneously, reducing them afresh with the software described by
\citet{ivison06}, though without the shift-and-add technique since
4C\,60.07 was not detected at high signal-to-noise (S/N) in individual
scans. The resulting images, smoothed with 7-arcsec Gaussians to yield
point spread functions with {\sc fwhm} of $\sim$10 and
$\sim$15\,arcsec at 450 and 850\,$\mu$m, are shown in
Fig.~\ref{scuba}. No positional corrections have been applied. Another
faint SMG\footnote{Labelled `1' in \citet{stevens03}, which contains
  spurious coordinates. The object should be referred to as
  SMM\,J051254.2+603119. It has no counterpart at 24\,$\mu$m and may
  lie at a similarly high redshift to 4C\,60.07.}  lies to the north,
with its reference beams visible, well outside the primary beam of the
SMA's 6-m antennas at 345\,GHz (30\,arcsec, {\sc fwhm}).

\section{Results}
\label{discussion}

\subsection{Submm morphology and flux density}
\label{morphology}

\begin{table}
\begin{center}
\caption{Submm and radio positions in the 4C\,60.07 system.}
\begin{tabular}{lcc} \\ \hline
Characteristic  & Value & Comment\\ \hline
R.A.\ (J2000)&05:12:55.13$\,\pm\,$0.04&NE SMA component `A'\\
Dec.\ (J2000)&+60:30:49.8$\,\pm\,$0.3  &\\
 ~~Peak$^a$ $S_{\rm 890\mu m}$  &$6.34\pm 1.36$\,mJy&\\
\smallskip
 ~~Total$^a$ $S_{\rm 890\mu m}$ &$8.72\pm 2.93$\,mJy&\\ 
R.A.\ (J2000)&05:12:54.71$\,\pm\,$0.03&SW SMA component `B'\\
Dec.\ (J2000)&+60:30:48.8$\,\pm\,$0.2  &\\
 ~~Peak$^a$ $S_{\rm 890\mu m}$  &$7.86\pm 1.39$\,mJy&\\
\smallskip
 ~~Total$^a$ $S_{\rm 890\mu m}$ &$7.83\pm 2.40$\,mJy&\\
R.A.\ (J2000)&05:12:55.147$\,\pm\,$0.003&Radio core, 4.7\,GHz\\
Dec.\ (J2000)&+60:30:51.01$\,\pm\,$0.02  &\\
R.A.\ (J2000)&05:12:55.154$\,\pm\,$0.004&Radio core, 8.2\,GHz\\
Dec.\ (J2000)&+60:30:51.03$\,\pm\,$0.03  &\\ \hline
\end{tabular}
\label{res-tab}
\end{center}

\noindent
$^a$Measured using a twin-Gaussian fit to the dirty image.
\end{table}

\begin{figure}
\begin{center}
\includegraphics[scale=0.45]{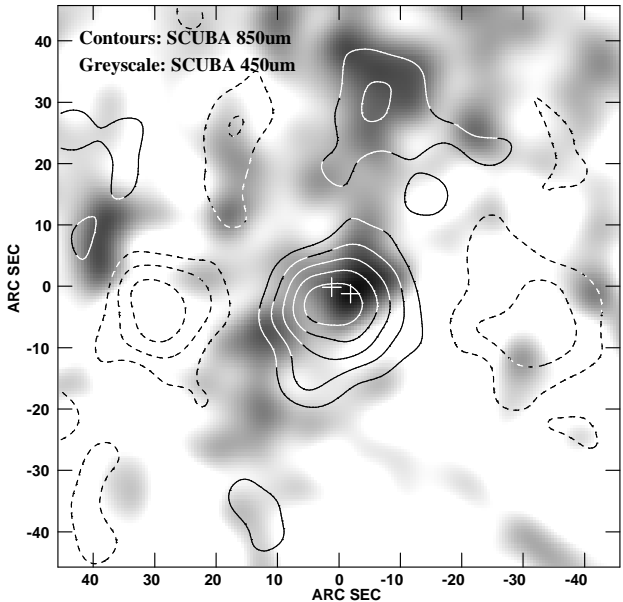}
\caption{Greyscale representation of 4C\,60.07 observed using SCUBA
(\S\ref{submm}) at 450\,$\mu$m, superimposed with contours of the
850-$\mu$m image (shown in its original form by \citealt{stevens03})
plotted at $-6, -4, -2, 2, 4, 6, 8, 10 \times\sigma$, where $\sigma$
is the noise level. The reference beams can be seen, due to the
east-west chopping and nodding of the secondary mirror on the
JCMT. The positions of the components detected by our SMA imaging,
discussed later, are shown as crosses.}
\label{scuba}
\end{center}
\end{figure}

\begin{figure}
\begin{center}
\includegraphics[scale=0.45]{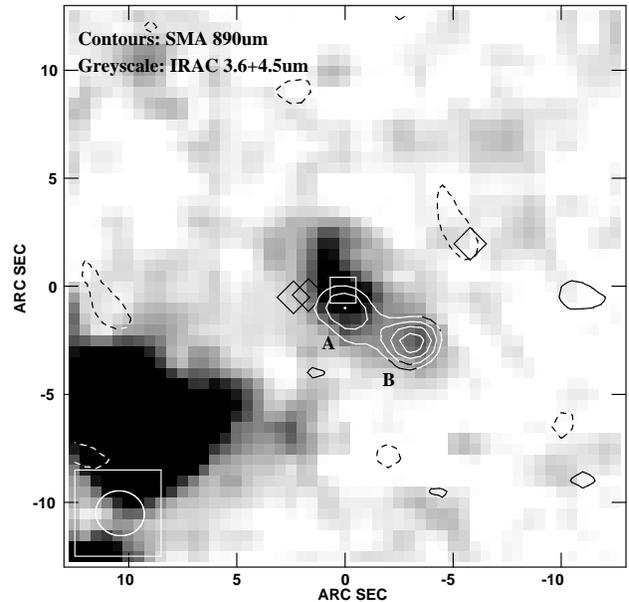}
\caption{Greyscale representation of the {\em Spitzer} averaged 3.6-
and 4.5-$\mu$m images of 4C\,60.07. Superimposed on the IRAC data is
the high-resolution SMA 890-$\mu$m image of 4C\,60.07, with contours
plotted at $-2.5$, 2.5, 3.5, 4.5, 5.5 and 6.5 $\times$
1.1\,mJy\,beam$^{-1}$, the r.m.s.\ noise level. The size and shape of
the SMA's synthesised beam is shown in the bottom left corner of this
plot. The {\sc fwhm} of the SMA's primary beam extends slightly beyond
the extent of the image, to $\sim$30\,arcsec {\sc fwhm}. The position
of the radio core -- as seen at 4.7 and 8.2\,GHz -- is shown by a
square; those of the steep-spectrum hot spots are labelled with
diamonds; the submm components, `A' and `B', are labelled. The IR,
radio and submm images are aligned to better than 0.5\,arcsec,
conservatively.}
\label{sma+irac12}
\end{center}
\end{figure}

As stated earlier, 4C\,60.07 is extremely bright at submm wavelengths,
with $S_{\rm 850\mu m}=23.8\pm 3.5$\,mJy, as measured via
fully-sampled jiggle maps with SCUBA (Fig.~\ref{scuba}). Using SCUBA's
photometry mode, which was insensitive to emission larger than the
telescope beam, the measured flux density was $17.11\pm
1.33$\,mJy\,beam$^{-1}$ -- the brightest of 47 radio galaxies observed
by \citet{archibald01} and \citet{reuland04}.

Our new, high-resolution submm interferometry, shown in
Fig.~\ref{sma+irac12}, reveals unambiguously that the emission first
detected using SCUBA is spread over a region larger than the
$\ls$500\,pc ($\ls$0.1\,arcsec at $z=3.8$) usually encountered in
local ULIRGs \citep{ds98}, or even the $\ls$4\,kpc ($\ls$0.5\,arcsec)
usually encountered in SMGs (e.g.\ \citealt{younger07, younger08,
bi08}, cf.\ SMM\,J094303+4700 and SMM\,J123707+6214 in
\citealt{tacconi06, tacconi08}). Two submm components are revealed,
hereafter `A' and `B'. Component `A' is the easternmost and
northernmost (see Table~\ref{res-tab}), lying 1.2\,arcsec south of the
flat-spectrum radio core that has hitherto been presumed to host the
AGN driving 4C\,60.07's prodigious radio luminosity. `A' and `B' lie
3.3\,arcsec apart (24\,kpc in the plane of the sky) and have similar
integrated flux densities ($S_{\rm 890\mu m}=8.7$ and
$7.8$\,mJy). Hereafter, `A', `B' and the radio core will be the
principal components discussed.

Component `A' is marginally resolved by our $\sim$2-arcsec synthesised
beam, with a size reported by {\sc jmfit} in \AIPS\ of [2.2 $\times$
0.4]\,arcsec at PA = 78$^{\circ}$; its peak $S_{\rm 890\mu m}$ is
lower than that of `B', but its integrated $S_{\rm 890\mu m}$ is
marginally higher. The total $S_{\rm 890\mu m}$ of these two submm
components, $16.6\pm 3.8$\,mJy, the equivalent of $18.7\pm 4.3$\,mJy
at 850\,$\mu$m (applying a +13-per-cent correction for a 55-{\sc k}
greybody with emissivity index, +1.5). This is consistent with the
$23.8\pm 3.5$\,mJy measured by SCUBA at 850\,$\mu$m at the
$\sim$1-$\sigma$ level.

\begin{figure}
\begin{center}
\includegraphics[scale=0.45]{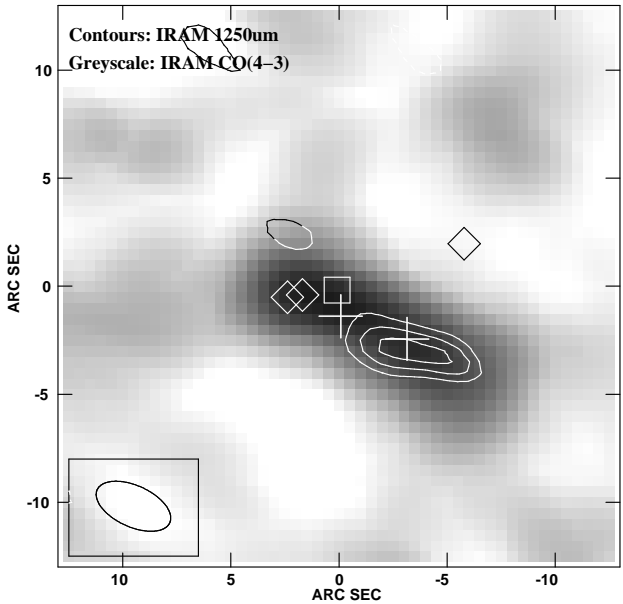}
\caption{Greyscale representation of the uniform-weighted,
velocity-integrated CO(4--3) emission from 4C\,60.07, as detected
using IRAM PdBI with a [8.9 $\times$ 5.5]-arcsec synthesised beam by
\citet{papadopoulos00}, superimposed with contours of their 240-GHz
(1.25-mm) continuum image, plotted at $-2.5$, 2.5, 3.5, 4.5, 5.5 and
6.5 $\times$ 0.58\,mJy\,beam$^{-1}$, the r.m.s.\ noise level, with the
shape and size of the somewhat elongated 240-GHz synthesised beam
shown in the bottom left corner. The image is centred on the position
of the radio core -- seen at 4.7 and 8.2\,GHz -- which is labelled
with a square; those of the steep-spectrum hot spots are labelled with
diamonds; submm components, `A' and `B' (see Fig.~\ref{sma+irac12}),
are marked with crosses.}
\label{co}
\end{center}
\end{figure}

The submm morphology appears consistent with that seen at lower S/N
using the Plateau de Bure Interferometer (PdBI) of the Institut de
Radio Astronomie Millim\'etrique (IRAM) at 240\,GHz (Fig.~\ref{co} --
\citealt{papadopoulos00}) and confirms the extended nature of the dust
emission. The measured size, even taking both components together, is
significantly smaller than the [11 $\times$ 4]-arcsec deconvolved size
estimated by \citet{stevens03}, and the single-dish and
interferometric flux densities would allow up to 5\,mJy of emission
distributed on scales larger than those probed by the SMA. Overall,
our observations show little compelling evidence for the galaxy-wide
starburst reported by \citet{papadopoulos00} or \citet{stevens03}.

\begin{figure}
\begin{center}
\includegraphics[scale=0.45]{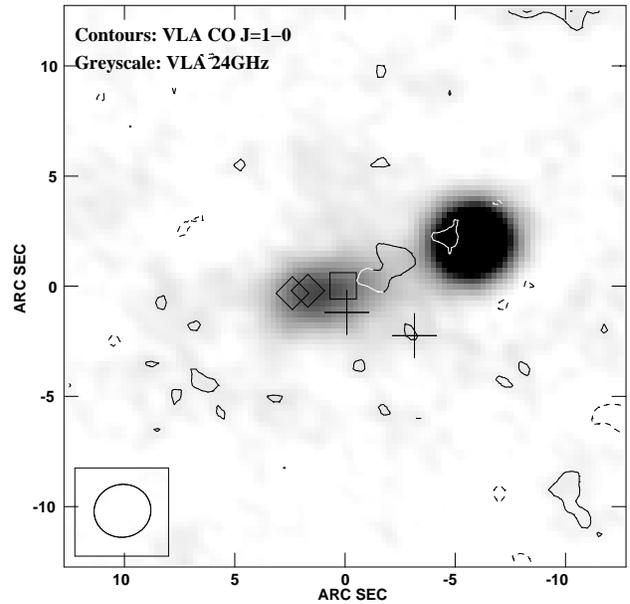}
\caption{Greyscale representation of the 24-GHz continuum emission
from 4C\,60.07 superimposed with contours of the CO(1--0) emission in
the velocity range of the broad CO(4--3) component detected with IRAM
PdBI by \citeauthor{papadopoulos00} ($+455\pm312$\,km\,s$^{-1}$) as
imaged using the VLA with a [2.6 $\times$ 2.4]-arcsec synthesised beam
(shown, bottom left). The CO(1--0) data shown are the average of the
first two methods described in \S\ref{radio}. Contours are plotted at
$-2$, 2 and 3 $\times$ 18.4\,$\mu$Jy\,beam$^{-1}$, the r.m.s.\ noise
level. The emission coincident with component `B' has an overall
significance, using a Gaussian fit with size fixed to that of the
synthesised beam, of $\ls$2\,$\sigma$. The image is centred on the
position of the radio core, which is labelled with a square; those of
the steep-spectrum hot spots are labelled with diamonds; submm
components, `A' and `B' (see Fig.~\ref{sma+irac12}), are marked with
crosses.}
\label{co1-0}
\end{center}
\end{figure}

\subsection{CO morphology}
\label{co-morph}

\citet{papadopoulos00} found the CO $J=4-3$ line emission from
4C\,60.07 to be bright and broad, even by the standards now known for
SMGs \citep{neri03, greve05}. They describe two distinct CO(4--3)
components in their relatively low-spatial-resolution PdBI data cube
(Fig.~\ref{co}). Together, these cover $\gs$1,000\,km\,s$^{-1}$.

One CO(4--3) component is spans only $\sim$150\,km\,s$^{-1}$ {\sc
fwhm}. Its position is consistent with that of the putative AGN core
-- taken to be the flat-spectrum radio source between the
steeper-spectrum hot spots seen most clearly at 1.4\,GHz -- and it is
$\sim$220\,km\,s$^{-1}$ blueward of the systemic (He\,{\sc ii})
velocity of the AGN's host galaxy. \citet{gip04} detected gas glowing
in CO(1--0), coincident in velocity with the narrow CO(4--3) emission,
and consistent spatially. On the basis of dynamical mass limits,
\citeauthor{papadopoulos00} argued that this narrow component has yet
to form the bulk of its eventual stellar mass.

The second CO(4--3) component lies $\sim$700\,km\,s$^{-1}$ redward and
$\sim$7\,arcsec south-west of the narrow component and spans
$\gs$550\,km\,s$^{-1}$ {\sc fwhm}, with emission likely extending to
lower frequencies, beyond the observed band \citep{papadopoulos00}.
\citet{gip04} report CO(1--0) in the velocity range of this broad
CO(4--3) component, with a peak flux density of $0.27\pm 0.05$\,mJy,
but offset spatially by $\sim$4\,arcsec to the north-east, lying just
west of the radio core.

However, our re-analysis of the data used by \citeauthor{gip04} to
explore the broad CO component, together with a much larger set of
data taken with shorter baselines, have revealed no CO(1--0) emission
above 0.05\,mJy\,beam$^{-1}$ (Fig.~\ref{co1-0}). The total CO(1--0)
flux, integrated over the velocity range to which these data are
sensitive ($+455\pm312$\,km\,s$^{-1}$), is $3\sigma <
0.07$\,Jy\,km\,s$^{-1}$. Adopting $X_{\rm CO} = 0.8$\,({\sc
k}\,km\,s$^{-1}$\,pc$^{-1}$)$^{-1}$\,M$_{\odot}$ \citep{ds98}, this
translates into $M_{\rm H_2} < 3.6\times 10^{10}$\,M$_{\odot}$, less
than half of the mass determined on the basis of the CO $J=4-3$
emission by \citeauthor{papadopoulos00} who assumed a 4--3/1--0 line
ratio of $r_{43}=0.45$, and less than the mass determined for the
narrow CO(1--0) component seen blueward of our passband by
\citet{gip04}. This suggests variations in the bulk gas properties
across the 4C\,60.07 system.

\subsection{Comparison of IR, submm and radio positions}
\label{positions}

\begin{figure*}
\begin{center}
\includegraphics[scale=0.47]{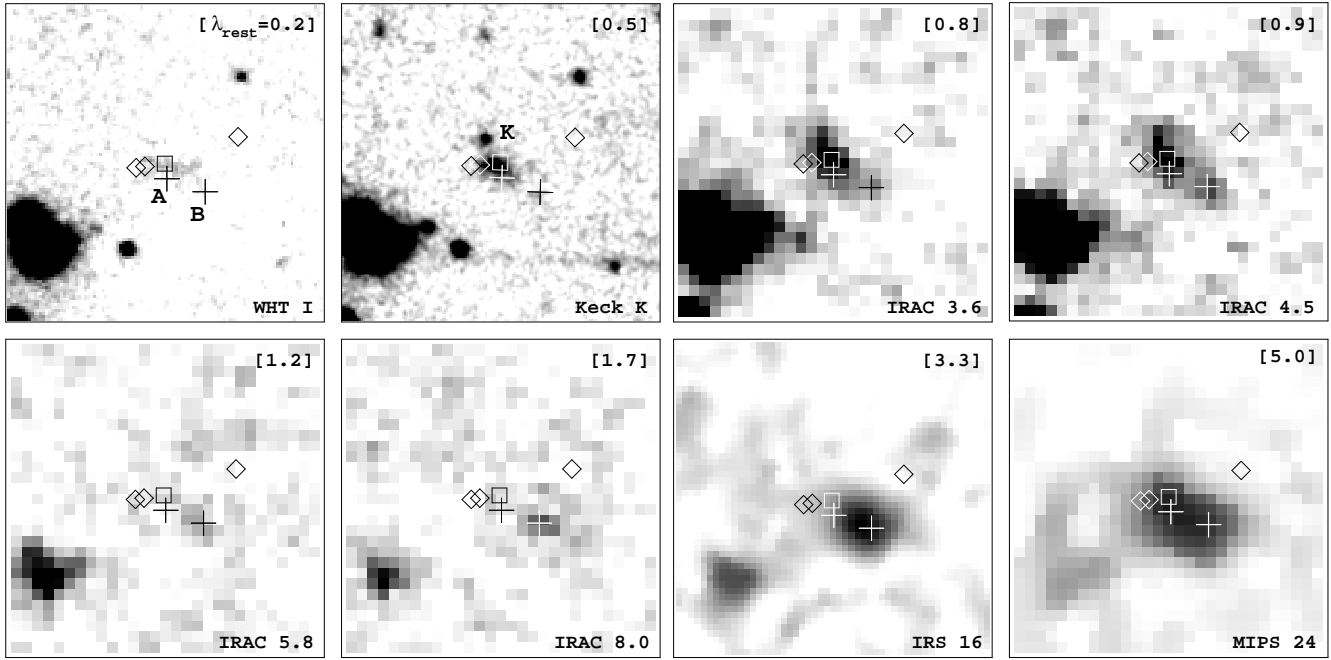}
\caption{Greyscale representations of WHT $I$ (\S\ref{optir}), Keck
$K'$ \citep{vb98}, and {\em Spitzer} images of 4C\,60.07 from 0.9
through to 24\,$\mu$m, using a linear stretch from 0--10\,$\sigma$,
where $\sigma$ is the pixel noise (0--20\,$\sigma$ at 24\,$\mu$m,
where we have re-mapped the image into 0.86-arcsec pixels to match the
IRAC images, 2.9$\times$ smaller linearly than is necessary to fully
sample the 24-$\mu$m point spread function). The WHT and Keck images
have been smoothed with a 0.3-arcsec Gaussian. The Keck image has been
rotated slightly to match the USNO-based astrometry of the WHT
$I$-band image. In each image, two crosses show the positions of the
SMA submm peaks and the approximate rest-frame wavelength (in $\mu$m)
is shown in parentheses. The images are centred on the position of the
radio core, labelled with a square, while those of the steep-spectrum
hot spots are labelled with diamonds. Each image is
[25\,$\times$\,25]\,arcsec.}
\label{mosaic}
\end{center}
\end{figure*}

We first concern ourselves with the new data from the SMA. The most
unexpected aspect of the SMA, {\em Spitzer} and VLA imaging
(Fig.~\ref{sma+irac12}) is the offset between the primary sites of
submm emission and that of the radio core's synchrotron\footnote{The
positional offset is significant at the 4-$\sigma$ level. The
uncertainty in the submm position is dominated by the centroiding
\citep[see the appendix of][]{ivison07} since systematic uncertainties
are limited by the coincidence of the southern submm component with IR
emission and by the positional tests described in \S\ref{sma}.}. The
submm components lie 10--30\,kpc (on the plane of the sky) from the
radio core -- a larger distance than can be understood via the
positional uncertainties ($\ll$4\,kpc; see Table~\ref{res-tab} and
\S\ref{observations}). The large offsets often seen between
single-dish submm detections and their radio or IR counterparts
\citep{ivison07} are not expected here, due to the small SMA
synthesised beam and accurate phase-referenced interferometric
positions. It is clear, therefore, that the radio galaxy is not
associated directly with rest-frame far-IR emission\footnote{Of the
six submm-bright HzRGs with deep {\em Spitzer} imaging
\citep{archibald01, reuland04, seymour07}, all display an accurate
alignment of the radio core and the 3.6-$\mu$m emission (to a
tolerance of $\sim$1\,arcsec) and only 4C\,60.07 shows a misalignment
at 8\,$\mu$m (Fig.~\ref{mosaic}).}.

Fig.~\ref{mosaic} shows how the morphology of the 4C\,60.07 system
changes as we move from rest-frame 0.2\,$\mu$m (UV), via rest-frame
0.5\,$\mu$m ($\sim V$), 0.8\,$\mu$m ($\sim R$), 0.9\,$\mu$m ($\sim
I$), 1.2\,$\mu$m ($\sim J$), 1.7\,$\mu$m ($\sim H$) and 3.3\,$\mu$m
($\sim L$) to rest-frame 5.0\,$\mu$m ($\sim M$), i.e.\ $\lambda_{\rm
obs}$ = 0.9--24\,$\mu$m. The rest-frame UV emission is centred on the
radio core. In rest-frame $V$, a red object (`K') becomes visible,
2.2\,arcsec NNE of the core; the core emission covers $\sim$2\,arcsec
and there is only low-level flux from the submm components. `K' makes
contributions in rest-frame $R$ and $I$ ($\lambda_{\rm obs}$ =
3.6--4.5\,$\mu$m) and the emission coincident with submm component `B'
strengthens. By the time we reach the rest-frame $J$ and $H$ filters
at $\lambda_{\rm obs}$ = 5.8--8.0\,$\mu$m, `B' has become the dominant
component, with only very weak emission associated with the radio core
and `K'. Component `A' appears to be devoid of significant mid-IR
emission, with less than 4, 6, 26 and 39\,$\mu$Jy (3\,$\sigma$) at
$\lambda_{\rm obs}$ = 3.6, 4.5, 5.8 and 8.0\,$\mu$m. At $\lambda_{\rm
obs}$ = 24\,$\mu$m we see emission along the orientation of the submm
components, `A' and `B', with little room for a significant
contribution from the radio core (Fig.~\ref{mosaic}). (The MIPS
24-$\mu$m filter transmits between 4.3 and 5.4\,$\mu$m -- rest frame,
to half power -- and so misses the strongest PAH features, e.g.\
\citealt{desai07}).

\subsection{Clues from SEDs via IR diagnostic plots}
\label{diagnostics}

\begin{table}
\begin{center}
\caption{Mid-IR flux densities.}
\begin{tabular}{lcc} \\ \hline
Isophotal wavelength& \multicolumn{2}{c}{Flux density ($\mu$Jy)}\\
($\mu$m)            &Core/`A'            & `B'\\ \hline
3.535               &20.6\,$\pm$\,1.4    &5.9\,$\pm$\,1.3\\
4.502               &27.0\,$\pm$\,2.0    &12.1\,$\pm$\,2.0\\
5.65                &3$\sigma$\,$<$\,26.0&29.0\,$\pm$\,9.0\\
7.74                &3$\sigma$\,$<$\,39.0&76.0\,$\pm$\,13.0\\  \hline
16.0&\multicolumn{2}{c}{175\,$\pm$\,33}\\
24.0&\multicolumn{2}{c}{930\,$\pm$\,100}\\ \hline
\end{tabular}
\label{fluxes-tab}
\end{center}
\end{table}

\begin{figure}
\begin{center}
\includegraphics[scale=0.47,angle=270]{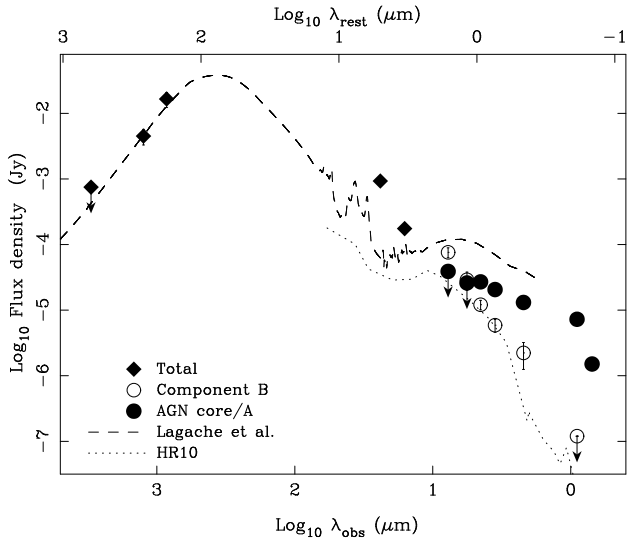}
\caption{SEDs of the two main components of 4C\,60.07 from the optical
to the submm, with component `B' represented by open circles and the
AGN core by filled circles. Beyond 8\,$\mu$m we show total flux
densities, as diamonds. The dotted line is the SED of the $z=1.44$
ERO, HR\,10 \citep{err88, stern06}, normalised in flux with respect to
4C\,60.07 at rest-frame 185\,$\mu$m (only the well-sampled optical/IR
SED is plotted); the dashed line is the SED of a model
10$^{13}$-L$_{\odot}$ starburst \citep{lagache03}.}
\label{sed}
\end{center}
\end{figure}

We have extracted $I$ and 3.6--8.0-$\mu$m flux densities for the two
components, the radio core and `B', that dominate the rest-frame
optical--IR emission from the 4C\,60.07 system, and adopted other
measurements from \citet{chambers96} and \citet{vb98}. The different
noise levels at 3.6--8.0\,$\mu$m in Fig.~\ref{mosaic} make it
difficult to judge the spectral energy distribution (SED) shapes, so
we plot both of the dominant components in Fig.~\ref{sed} and list the
flux densities in Table~\ref{fluxes-tab}. The rest-frame UV--$V$
emission from the core is resolved. At rest-frame $R$ and $I$
($\lambda_{\rm obs}=3.6-4.5$\,$\mu$m) we see a blend of emission from
the core and `K'; since our photometry is appropriate for point
sources, the true flux densities are probably $\sim$20 per cent higher
than those shown. Component `B' is barely resolved (3.1\,arcsec, {\sc
fwhm}), so its photometry ought to be accurate. The 24-$\mu$m emission
is a blend, with component `B' contributing less than half. The
total\footnote{\citet{seymour07} report different 8- and 24-$\mu$m
flux densities for 4C\,60.07. Their 7-arcsec aperture, centred on the
3.6-$\mu$m emission, misses around half the 8-$\mu$m flux (which comes
mostly from `B'). At 24\,$\mu$m, \citeauthor{seymour07} used an
aperture 26\,arcsec in diameter (not 13\,arcsec as stated in the
paper; N.\ Seymour 2008, private communication) and thus include flux
from the source to the south-east. Their 16-$\mu$m aperture suffers no
such problem, although it was 12\,arcsec in diameter rather than
6\,arcsec, as stated, so we adopt their 16-$\mu$m flux here.} flux
density at 24\,$\mu$m is $0.97\pm0.05$\,mJy.

AGN SEDs rise as a power law between 1 and 10\,$\mu$m (rest frame),
while starbursts have flatter SEDs at rest-frame 1--3\,$\mu$m which
rise at longer wavelengths \citep[e.g.][]{ward87,donley07}. At
$z=3.8$, IRAC probes rest-frame 0.8--1.7\,$\mu$m, so $S_{\rm 8.0\mu
m}/S_{\rm 4.5\mu m}$ is a powerful AGN diagnostic, particularly when
plotted against a probe of the $>$3\,$\mu$m slope, such as $S_{\rm
24\mu m}/S_{\rm 8\mu m}$ \citep{ivison04, lacy04, magdis08}.

If the 3.6--5.8-$\mu$m emission associated with the radio galaxy core
is associated with starlight then we might expect that emission to
continue through to $\lambda_{\rm obs}$ = 7.7\,$\mu$m -- near the
isophotal wavelength of the IRAC 8.0-$\mu$m filter, which corresponds
to the 1.6-$\mu$m stellar bump at $z\sim3.8$.  We do not see this, but
using the IRAC photometry to constrain rest-frame $H$-band
luminosities and assuming $L_H/M\sim 2$, the stellar mass is poorly
constrained: $<10^{12}$\,M$_{\odot}$, cf.\ $(2.8\pm 1.2)\times
10^{11}$\,M$_{\odot}$ in \citet{seymour07}, who found the $H$-band
light to be 95-per-cent stellar. We note that a much larger $L/M$ (and
hence lower stellar mass) is plausible, via fine-tuning of age and
initial mass function, and there is probably a contribution to the
rest-frame near-IR light from the AGN. Our limit is therefore likely
to be very conservative, but is consistent with a large mass of
$\sim$0.2--1-Gyr-old stars in the radio galaxy host according to a
simple Starburst99 model \citep{vazquez05}. Already revealed as an AGN
by its radio emission, the core does not have the optical--IR
continuum characteristics of an AGN.

In contrast, the SED of component `B' climbs steeply to $\lambda_{\rm
obs}\sim 8.0$\,$\mu$m. It is fit well by a $S_{\nu}\propto \nu^{-3.3}$
power law: an extremely red galaxy, akin to HR\,10 (= ERO
J164502+4626.4) at $z\sim 1.4$ which has excellent coverage of its SED
-- see Fig.~\ref{sed} \citep{dey99, stern06}. The value of $S_{\rm
8.0\mu m}/S_{\rm 4.5\mu m}$ seen for component `B' ($6.3\pm1.5$, or
[4.5] $-$ [8.0] = 2.0) is unusual: redder than any SMG in the SHADES
sample \citep{ivison07}, and ranking with the reddest of the
8-$\mu$m-selected objects explored by \citet{magdis08}. These
diagnostics are typically used for objects at $z\sim 2$, however, so
we should explore whether the colour remains extreme at $z\sim 4$.
Does it exhibit similar rest-frame optical/near-IR colours to HR\,10?
For $z=1.4$, $S_{\rm 3.6\mu m}/S_{\rm 2.2\mu m}$ is close to $S_{\rm
8.0\mu m}/S_{\rm 4.5\mu m}$ at $z=3.8$. HR\,10 has $S_{\rm 3.6\mu
m}/S_{\rm 2.2\mu m}\sim 2.35\pm0.24$, so although HR\,10 is redder
than component `B' at rest-frame $\ls$1\,$\mu$m, it is not nearly as
red at rest-frame 1--10\,$\mu$m which means that component `B' can be
categorised as an obscured AGN with little room for ambiguity.

\section{A toy model of the 4C\,60.07 system}
\label{model}

\begin{figure}
\begin{center}
\includegraphics[scale=0.42,angle=0]{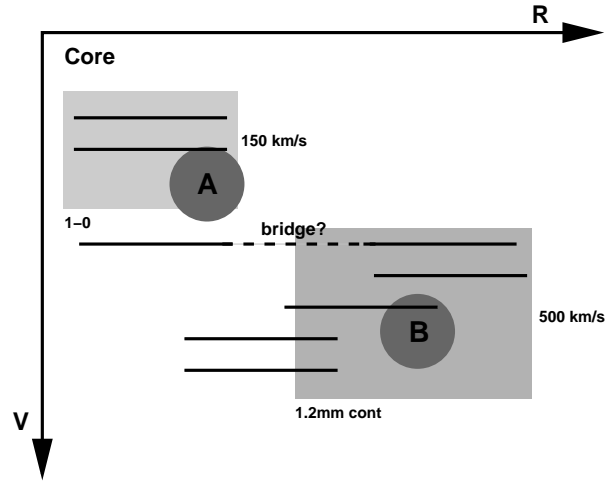}
\caption{Schematic of 4C\,60.07. The horizontal axis is a spatial
vector on the sky, starting at the radio core and running through `A'
and `B'. The vertical axis shows relative velocity, where
available. Black lines shown the extent of CO(4--3) channel maps from
\citet{papadopoulos00} -- the dashed line represents the emission that
spans both components (note the velocity gap between it and the
150-km\,s$^{-1}$ component). Components `A' and `B' are marked by
circles at arbitrary velocities. The upper box represents the CO(1--0)
map of the narrow component \citep{gip04}, while the lower box
represents the 1.2-mm continuum, its velocity chosen to match that of
the broad component.}
\label{cartoon}
\end{center}
\end{figure}

It is tempting to invoke a complex, many-component system and to
speculate that this form of interaction may be related to radio
loudness and the formation of the most massive galaxies, with
implications for the form and role of feedback. However, we must not
forget the origins of our target: 4C\,60.07 was selected initially as
a USS emitter and its radio activity is thus inescapably young,
$\le$10\,Myr \citep{br99}.  We are therefore seeing a system that is
probably out of equilibrium.

Can we marry together all the data for the 4C\,60.07 system -- in
particular the radio and submm interferometry, both continuum and
spectral line (CO), and the mid-IR imaging? These characteristics are
presented schematically in Fig.~\ref{cartoon}.

In our favoured model, 4C\,60.07 comprises a far-IR-luminous starburst
triggered by a galaxy-galaxy interaction. That more than one component
is involved is inescapable, given the morphologies evident in the
submm and mid-IR wavebands.  Of the myriad components in the system,
we identify the radio core as one with significant mass, since it
appears to comprise a relatively large BH (capable of driving the
radio lobes) and a host galaxy that is visible in rest-frame UV, $V$,
$R$ and $I$. Since it lacks rest-frame far-IR emission, we suspect the
radio galaxy host has exhausted (or driven away) its supply of
molecular gas. Certainly, little H$_2$ is evident via its most
sensitive available tracers, dust and CO. We identify component `B' as
the other high-mass component, as evident from its very broad CO(4--3)
line emission -- a dusty, gas-rich galaxy, as evident from the submm
emission, undergoing a huge burst of star formation. Component `B' is
very heavily enshrouded in the dust that betrays its presence in the
submm waveband, visible elsewhere in its SED only longward of
rest-frame $R$, probably via a buried AGN with a power-law SED
(\S\ref{diagnostics}).

What of component `A'? The key piece of evidence is its narrow CO line
width, seen in both $J=4-3$ and $1-0$. \citet{papadopoulos00} believed
this emission was associated with the radio galaxy host but our
high-resolution SMA imaging, together with the high-resolution
CO(1--0) observations of \citet{gip04}, betray an association with
component `A'.  We believe that `A' is a bridge of cold material
tidally stripped from the massive, gas-rich starburst/AGN (`B')
towards the host of the radio core -- an analogue of the scenario seen
in some local ULIRGs, but on a more massive scale. This is in accord
with its observed characteristics: the submm component has no
counterpart at mid-IR wavelengths, is resolved on a scale of
$\sim$2\,arcsec ($\sim$15\,kpc) and its gas -- consistent spatially
for CO(1--0) and CO(4--3) -- has a very low velocity dispersion
($\sim$150\,km\,s$^{-1}$).

\section{Concluding remarks}
\label{conclusions}

New, high-resolution submm imaging of the HzRG, 4C\,60.07, at $z=3.8$,
reveals two clumps of emission separated by 3.3\,arcsec
($\sim$25\,kpc). Although one clump is resolved by our $\sim$2-arcsec
beam, both are relatively compact ($\ls$15\,kpc) and there is little
evidence for a galaxy-wide starburst on larger scales from our
observations.

Previously thought to comprise two gas-rich starbursts, one centred on
the radio galaxy, we discover an extra level of complexity due to the
misalignment of the submm emission and the radio core.  Given the
extreme youth of HzRGs, we interpret the 4C\,60.07 system as an
early-stage merger, with its radio jets and a nearby gas-rich
starburst/AGN both induced by the interaction. We interpret a second
clump of dust, resolved by our submm imaging and coincident with very
narrow CO emission, as a plume of cold, dusty gas in a tidal stream,
seen in a non-equilibrium state. It is separated spatially from the
radio galaxy, which may have exhausted its supply of fuel.

Recent work on the $z=4.7$ quasar BR\,1202$-$0725 \citep{feain04}, on
LH\,850.7 \citep{ivison02} and on Minkowski's Object \citep{croft06}
has suggested that some starbursts are induced by the interaction of a
radio galaxy jet with a large gas reservoir. The story we are seeing
unfold in the 4C\,60.07 system raises the possibility that some of
these jets may have been triggered by the interaction that led to the
starburst, rather than the starburst being triggered by the jet.

When a BH is spatially offset from the gas and dust of its host
system, as seen in 4C\,60.07, this has implications for the
calculation of its dynamical mass (via the extent of its CO, spatially
and in velocity) and related claims about the co-assembly of BHs and
host spheroids at high redshift \citep[e.g.][]{walter03, walter04},
regardless of whether the system in question is a disk or a merger.

Our observations of 4C\,60.07 bring to mind scaled-up versions of the
NGC\,4038/4039 (`Antennae' -- \citealt{wang04}) and VV\,114
\citep{frayer99} systems -- early-stage mergers where the bulk of the
star formation occurs in an obscured `overlap region' between two
gas-rich nuclei. In NGC\,4038/4039 this region displays a heavily
enhanced star-formation rate yet contains little stellar mass
\citep[$\sim$10 per cent of the total --][]{wang04}. In 4C\,60.07, the
galaxy nuclei correspond to the radio core and `B', and the overlap
region is `A'. There are also similarities between 4C\,60.07 and
HE\,0450$-$2958, one of the objects originally thought to have been
caught during its transition from a ULIRG into an optically bright
quasar \citep{hn88, sanders88, cs01} but now believed to be a gas-rich
spiral in collision with -- and stimulating optical activity in -- an
unobscured quasar hosted by a gas-poor elliptical
\citep{papadopoulos08}.

At $z\gs3$, where quasars are becoming rarer, a binary quasar with a
small separation ($<$1\,arcmin) has yet to be found \citep[at $2<z<3$,
tens are known --][]{mortlock99, kochanek99, hennawi06}. The 4C\,60.07
system contains two AGN and, although not strictly speaking a `binary
quasar', it does support the argument made by \citet{djorgovski91}
that the high number of binary quasars on small angular separations --
orders of magnitude more than that predicted by an extrapolation of
the quasar correlation function power law -- is due to the triggering
of AGN during mergers \citep[see also][]{smail03, alexander03, iono06}.

We would not have been aware of the BH that presumably drives the
immense radio luminosity of 4C\,60.07 were it not for the interaction
that triggered both the FR\,{\sc ii} radio activity and the nearby
starburst. Is there an unseen population of BHs at high redshift --
invisible unless actively fueled, yet sufficiently massive to drive a
$>$10$^{27}$-W radio source? Aside from its rare radio properties, the
SED of 4C\,60.07's host galaxy is not particularly unusual, and such
objects would be impossible to sift from the $\sim5\times 10^8$
similarly bright IRAC sources in the sky.

\citet{reuland07} claim that HzRGs and SMGs differ only in the
activity of their AGNs.  Why, then, are complex submm morphologies the
exception rather than the rule in typical SMGs? The answer lies in the
youth of the 4C\,60.07 system, as set by its selection as a USS radio
emitter. We see 4C\,60.07 in the first throes of a violent
interaction, probably within a few Myr of its latest burst of activity
commencing, whereas analysis of the rest-frame optical properties of
SMGs show that they are typically $\ge$10$\times$ older than HzRGs --
$\sim$100\,Myr \citep{smail04,swinbank06} -- and are therefore more
likely to display more relaxed morphologies. This provides a ready
explanation for the extended nature of the submm emission seen towards
the \citeauthor{stevens03} sample of HzRGs, which is at odds with the
compact emission seen for most SMGs \citep{younger07, younger08,
bi08}.  Since these HzRGs will all be youthful, we expect
non-equilibrium morphologies to be commonplace, though not necessarily
the complex dual-AGN signature seen in 4C\,60.07. It is not the
activity of their AGNs that mark HzRGs as different from SMGs, it is
their youth. The possibility that HzRGs differ from SMGs has many
implications. It would be desirable to make similar observations of a
larger sample of HzRGs at a resolution similar or better than that
employed here.

Recent studies of rest-frame far-IR emission, either directly
\citep{tacconi06, younger07, younger08} or via the far-IR/radio
correlation \citep{chapman04,bi08}, have highlighted the importance of
high spatial resolution. Single-dish observations -- with SCUBA2
\citep{holland06}, {\em Herschel} \citep{griffin07} or {\em SPICA}
\citep{swinyard08} -- will often provide an ambiguous or misleading
picture. Our work re-inforces this view: we have spent a decade
assuming that the radio core of 4C\,60.07 dominates the rest-frame
far-IR emission when the emission is, in fact, associated with a
gas-rich companion. We must await sensitive interferometers such as
the Atacama Large Millimetre Array \citep[ALMA --][]{wootten03} to
probe the interactions that drive many aspects of galaxy evolution,
and design future missions, such as the {\em Far-Infrared
Interferometer} \citep[{\em FIRI} --][]{hi08}, with the ability to
probe a large range of spatial scales.

\vspace*{5mm}
\noindent
{\it Facilities:} {SMA, \em Spitzer}, VLA, JCMT, WHT.

\bibliographystyle{mnras}
\bibliography{ivison}

\bsp

\end{document}